# High-throughput extraction of the anisotropic interdiffusion coefficients in hcp Mg-Al alloys

Jingya Wang[a,b], Weisen Zheng[c], Guanglong Xu[d], Javier Llorca[a,b], Yuwen Cui[d,e*]

[a] IMDEA Materials Institute. C/ Eric Kandel 2, 28906 Getafe, Madrid, Spain

[b] Department of Materials Science, Polytechnic University of Madrid/Universidad Politécnica de Madrid, E.T.S. de Ingenieros de Caminos, 28040 Madrid, Spain

[c] School of Materials Science and Engineering, Shanghai University, 200444 Shanghai, PR China

[d] Tech Institute of Advanced Materials & College of Materials Science and Engineering, Nanjing Tech University, 210009, Nanjing PR China

[e] ICMA, CSIC-Universidad de Zaragoza, 50009 Zaragoza, Spain

* Corresponding author, ycui@unizar.es

**Abstract**

A high-throughput experimental approach is presented to extract the anisotropic interdiffusion coefficient by combining information over the composition profiles obtained by the electron probe microanalysis (EPMA) and the grain orientation spectrum by the electron backscatter diffraction (EBSD) on polycrystalline diffusion couples. Following the forward-simulation scheme, the interdiffusion coefficients in grains with diverse orientation are obtained and subsequently used to determine the anisotropic interdiffusion coefficients perpendicular ($D_\perp(x)$) and parallel ($D_\parallel(x)$) to the *c* axis of the hcp Mg lattice in a Mg-Al alloy as a function of the Al solute content at 673 K and 723 K, respectively. It was found that the interdiffusion coefficients generally increased with the Al content and the rotation angle with respect to the *c* axis with a valley point around θ ≈ 30 ° at 723 K. And it was noticed that diffusion along the basal plane was always faster than along the *c* axis. A comprehensive explicit expression of the interdiffusion coefficients was provided as a function of Al content, grain orientation and temperature. The anisotropic impurity diffusion coefficients of Al in hcp Mg derived by the extrapolation of the results in this paper are in good agreement with first principles calculations in the literature.

# 1 Introduction

Mg alloys have received considerable attention in recent years due to their potential applications in the automotive, aerospace and biomedicine due to their low density, earth abundance, good castability, high specific stiffness and biocompatibility [1,2]. Al and Zn are often used as alloying elements to increase the strength of both cast and wrought Mg alloys. The microstructure and its evolution of these alloys during solidification and thermo-mechanical deformation are determined by the diffusion of the alloying elements through the anisotropic Mg crystals, which present a hexagonal close-packed (hcp) structure. Furthermore, the creep resistance is also intimately related to the mass transport via vacancy movement [3,4].

The majority of investigations on the diffusion behaviour in Mg alloys were performed in polycrystalline Mg or Mg alloys. The impurity diffusion coefficients were mainly measured using radioactive or stable tracers, and the concentration or activity profiles were obtained through residual activity, secondary ion mass spectrometry (SIMS) analysis or serial sectioning methods [5–19], whereas the interdiffusion processes were normally studied using the diffusion-couple technique [13]. Generally, the experimental diffusion coefficients obtained with these strategies are expected to be effective diffusion resulting from the contributions of mixed mechanisms, i.e. a combination of direct volume diffusion and grain boundary diffusion. However, the diffusion in the hcp materials is anisotropic in nature and depends on the grain orientations. This anisotropy is triggered by the hcp lattice, which leads to two different solute jumps, one within the basal plane (A jumps) and another between adjacent basal planes (B jumps) [20–22], as illustrated in Fig 1. The solute jumps within the basal plane are supposed to be identical in all directions. Hence, the diffusion coefficient tensor is determined by two different diffusion coefficients, i.e. $D_\perp$ resulting from the A jumps along the basal plane with a contribution from the B jumps and $D_\parallel$ (parallel to the $c$ axis) associated with the B jumps only.

The precise experimental investigation of the anisotropic diffusion coefficients requires the single crystal or bi-crystal samples, which are, however, hard to fabricate. The available experiments of the self-diffusion of pure Mg and the impurity diffusion of Al and Zn in Mg alloys [23–27] were widely confined to one or two crystallographic orientations, and thus were discrete and incomplete. The self-diffusion coefficient along the basal plane was determined using the radiotracer serial sectioning method in single crystal Mg and it was found to be higher within the basal plane than that of along the $c$- axis with a diffusion anisotropy ratios of $D_\perp/D_\parallel \sim 1.24$ at 848 K [23] and $\sim 1.21$ at 903 K [24]. These results, later confirmed by first principles calculations [28], showed that the self-diffusion anisotropy ratio varied from 2.29 at 300 K to 1.29 at 650 K. Regarding to the impurity anisotropic diffusion, the diffusion coefficients of Al and Zn in hcp-Mg were firstly extracted by Das et al. [25–27] using the diffusion couple technique in Mg single crystals. It was found that the Al diffusion within the basal plane was $\sim 1.33$ and $\sim 1.18$ times faster than that along the $c$ axis at 638 K and 693 K, respectively. This investigation was later extended to polycrystalline Mg diffusion couples to study the Al impurity diffusion in hcp Mg by varying with the diffusion direction that was described by the rotation angle θ from the basal plane [25]. This anisotropic impurity diffusion behavior was also investigated by the first-principles calculations [29], which indicated that Al impurity diffusion coefficient was higher along the basal plane than that along the $c$ axis, and the difference decreased as the temperature increased.

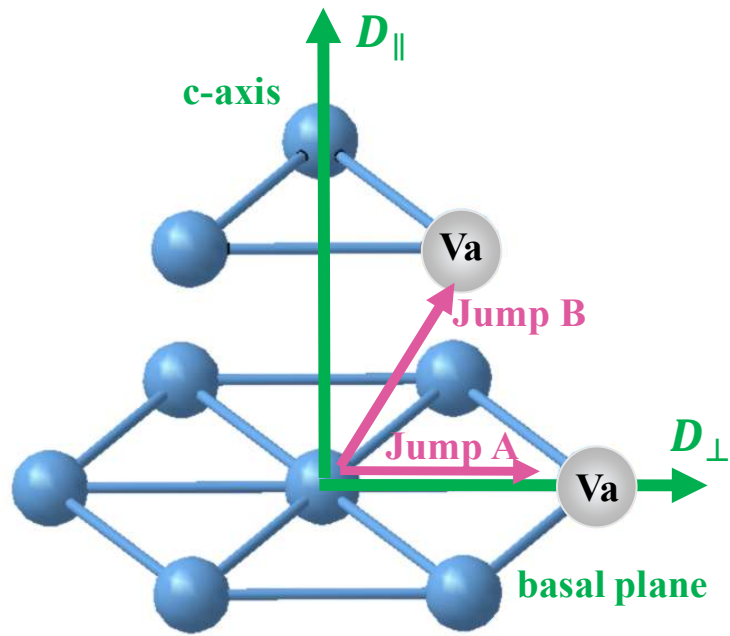

Fig. 1 Different diffusion jumps in a hcp lattice: jump A within the basal plane and jump B between adjacent basal planes. The diffusion coefficient $D_\perp$ depends on the jump A with partial contributions of the jump B, while only jump B contributes to the diffusion coefficient $D_\parallel$.

Current precise measurements of the anisotropic self-diffusion or impurity diffusion coefficients largely rely on the use of the radioactive and tracers, which is tedious and limited to a one-composition per measurement. More importantly, it is inapplicable to explore the anisotropic interdiffusion behavior or extract interdiffusion coefficients because of the technical difficulties associated with the introduction of the composition gradient. There are increasing efforts to utilize the single crystal or the bi-crystal end-member of the diffusion-couple to determine the anisotropic diffusion coefficients as a function of the grain orientation. Nevertheless, only the impurity diffusion information at the composition of the terminal end-member has been obtained so far, and the diffusion data are confined to one or two associated orientations. Meanwhile, the diffusion modeling remains challenging to study the complex interdiffusion behavior in binary or higher order hcp alloys. In consequence, the effect of grain orientation on the anisotropic interdiffusion in hcp Mg alloys is still missing and a robust approach is required to explore the anisotropic interdiffusion behavior and extract essential diffusion data as a function of the grain orientation.

This approach is presented here for a Mg-Al alloy. A polycrystalline diffusion couple containing multiple grains is used to explore the anisotropic interdiffusion behavior. The composition profiles are analyzed in the diffusion region across multiple large grains. The anisotropic diffusion coefficients are obtained as a function of the alloy composition and grain orientation using the forward-simulation scheme of multi-grain diffusion. This information is used to obtain the interdiffusion coefficients in the orientation perpendicular $D_\perp$ and parallel $D_\parallel$ to the $c$ axis as a function of Al solute content using the splitting technique developed in the present work. The paper is organized as follows: the experimental procedure is shown in Section 2, while Section 3 presents the approach to investigate the anisotropic diffusion behavior. The experimental results are analyzed and the anisotropic diffusion coefficients are obtained in Section 4. The main conclusions are finally summarized in Section 5.

## 2. Experimental procedures
### 2.1 Fabrication of the diffusion couple

The polycrystalline diffusion couple, Mg/Mg-9 at.%Al, was fabricated by using high-purity Mg (99.99 wt.%) and Al (99.99 wt.%). The pure Mg and Mg-9 at.%Al alloy were melted and cast in an induction furnace (VSG 002 DS, PVA TePla) under a protective Ar atmosphere to avoid the oxidation. During melting, the power increased at a rate of 0.1 kW/minute until 3.0 kW to ensure a steady heating process, and the melt was mixed thoroughly for five minutes before casting. The ingots were homogenized under an Ar atmosphere in quartz capsules at 673 K for 15 days to obtain a uniform composition and large grains (> 1 mm). The ingots were then cut into discs of 12 mm in diameter and 7.5 mm in length followed by the standard grinding process to 2000 grit and then polishing with 0.25 μm diamond paste to obtain a mirror-like surface. The Mg/Mg-9 at.%Al diffusion couples were prepared by diffusion-bonding the pure Mg and the alloy discs under a compressive load of 800 N for 1 h at 673 K in vacuum in a Gleeble 3800. The diffusion couples were annealed at 673 K (400 ℃) for 352 h and at 723 K (450 ℃) for 215 h in quartz capsules under the protection of Ar atmosphere, followed by quenching in water. The times and temperatures ensured that the diffusion couples reached a steady state and that the thermal stresses were fully released.

Several small pieces (7x7x2 mm$^3$) were cut along the direction perpendicular to the contact plane from the as-fabricated diffusion couples using a wire cutting machine at a low cutting speed to avoid the surface deformation. Subsequent grinding was manually performed using the abrasive SiC papers with the grit size of 320, 600, 1200 and 2000. Afterwards, the surface was manually polished with a conventional polishing machine to obtain a mirror-like flat surface using a MD-Mol cloth with 3 μm diamond paste, followed by a MD-Nap cloth with 0.25 μm diamond paste. The last step was chemical polishing with a solution of 75 ml ethylene glycol, 24 ml distilled water and 1 ml nitric acid to remove the surface deformation that resulted from the mechanical polishing process.

**2.2 Characterization**

The local compositions throughout the interdiffusion regions were measured by electron probe microanalysis (EPMA) using Wavelength Dispersive Spectroscopy (WDS) with a voltage of 20 KV, a beam current of 50 nA and the spot size of ~ 1 μm in the JEOL Superprobe JXA-8900M. The standard samples of high purity aluminum (99.99 wt.%) and high purity magnesium (99.99 wt.%) were used for the quantitative analysis of each element. In particular, the composition profiles were measured along lines parallel to the diffusion direction, which were perpendicular to the interface of the diffusion couples, as indicated in Fig. 2. Consequently, the microstructure was characterized by the scanning electron microscopy (SEM) and electron backscatter diffraction (EBSD) in a high-resolution scanning electron microscope (Helios Nanolab 600i FEI) equipped with an Oxford-HKL EBSD system. The EBSD measurements were conducted at the same locations where the diffusion profiles were measured using an accelerating voltage of 30 KV and a beam current of 2.7 nA with a step size of 1.5 μm. The EBSD maps provides the rotation angle, θ, that specifies the misorientation between the diffusion direction and the *c* axis of each grain (Fig. 2). Note that the rotation angle θ for each grain is directly obtained from the Euler angle provided by the EBSD measurements.

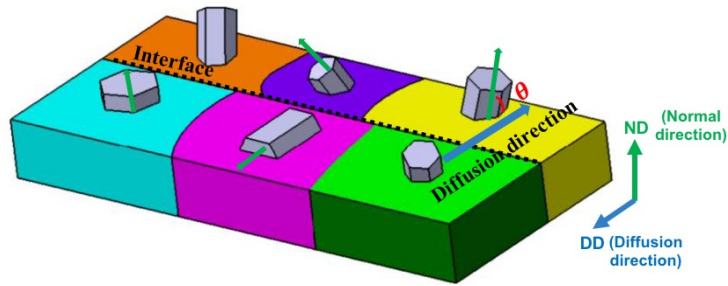

Fig. 2 Schematic representation of the rotation angle θ between the diffusion direction (DD) and the *c* axis of each grain.

## 3. Diffusion data analysis

The anisotropic interdiffusion coefficient $\widetilde{D}(x,\theta)$ within the interdiffusion region depends on both the grain orientation, indicated by the rotation angle ($\theta$), and the Al content ($x$). The analysis of the local composition profiles within the interdiffusion region in the diffusion couples can be used to extract the anisotropic interdiffusion coefficient $\widetilde{D}(x,\theta)$ for a large range of compositions and grain orientations. Afterwards, the experimental values of $\widetilde{D}(x,\theta)$ could be expressed as a function of two interdiffusion coefficients $D_\perp$ and $D_\parallel$ along the basal plane and parallel to the *c* axis of the Mg hcp lattice, respectively (Fig. 3).

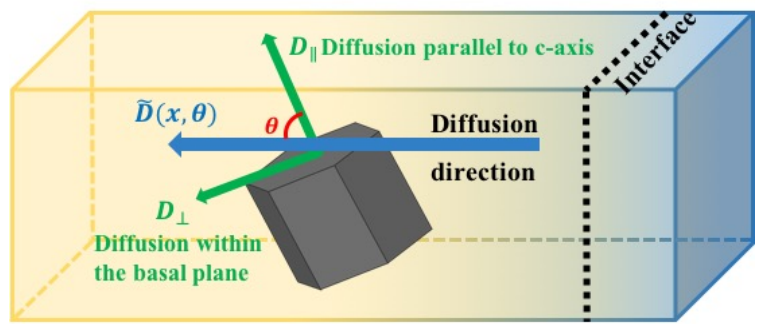

Fig. 3 Schematic representation of the anisotropic diffusion coefficients. The blue line stands for the diffusion direction, perpendicular to the interface indicated by the black dotted line, and the two components $D_\perp$ and $D_\parallel$ are represented by the green arrow line

### 3.1 Extracting interdiffusion coefficient with specific grain orientation

Eight specimens were sliced for analyzing the composition and characterizing the microstructure and grain orientation in this investigation, four of them were cut with great caution from the diffusion couples annealed at 673 K, and the other four were from those of 723 K. The microstructure and grain orientation information provided by EBSD measurement in the diffusion region of these eight specimens are shown in Fig. 4, where the grain color stands for the rotation angle $\theta$ between the $c$ axis of the grain and the diffusion direction. The composition profiles were measured by EPMA along the black lines (parallel to the diffusion direction) marked in each specimen, as depicted in Fig. 4. The left end of the diffusion couples corresponds to the pure Mg side while the right end stands for the Mg-Al alloy. It can be observed clearly that the interdiffusion region generally contains multiple large grains. It is hardly to obtain a diffusion couple that includes a single grain across the diffusion region due to the development of recrystallization during the high temperature annealing along the diffusion interface. These grains within the interdiffusion region are quite large (several hundreds of microns), and the measured composition profiles generally pass through multiple grains (Fig. 4). It should be noted the grain size in the Mg-rich region of the diffusion couple was sometimes quite small as ~ 50 μm (particularly in the samples annealed at 723 K), and the composition profile generally passed through multiple grains which were not considered in this investigation.

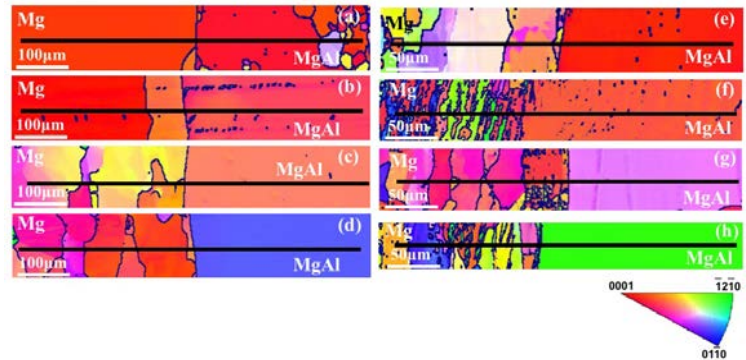

Fig. 4 Microstructure of the diffusion couples in the diffusion region. (a)-(d) Slice specimens annealed at 673 K. (e)-(h) Specimens annealed at 723 K. The color code stands for the grain orientation according to the inverse pole figure. The composition profile was measured along the black lines by EPMA.

The strategy to obtain the anisotropic diffusion coefficients from the composition profile is detailed below for the representative example as demonstrated in Fig. 5. In this case, the composition profile was obtained along the dark line parallel to the diffusion direction across three grains, Grain 1, 2 and 3, with the rotation angle of 5°, 29° and 33° respectively, as shown in Fig. 5a. And the composition profile to determine the interdiffusion coefficients spanned around 1 mm. The Al content along this line is plotted by the blue circles superposed to the microstructure in Fig. 5b, an enlarged square region in Fig. 5a. The raw measured composition profiles are analytically represented by the error function expansion (ERFEX) [30] as follows

$$x(\xi) = \sum_i a_i erfc[b_i \xi - c_i], \qquad (1)$$

where $x$ is the composition of Al expressed in at.%, $\xi$ is the Boltzmann transformation variable, $\xi = z/\sqrt{t}$, $z$ is the diffusion distance, $t$ is the diffusion time and $a_i$, $b_i$ and $c_i$ are a series of adjustable parameters to fit the experimental profile to the ERFEX function. The origin for the distance $z$ of the composition profile in Fig. 5b is the point for which the Al content begins to depart from 0. As stated previously [31,32], the ERFEX function provides a precise and robust representation of discrete experimental data obtained from the EPMA measurement, while eliminating the point-to-point composition fluctuations, yet allows more sound physical meaning by applying the error function to diffusion. This analytical representation of the diffusion profiles provided by ERFEX approximation is critical to attain reliable interdiffusion and impurity diffusion coefficients.

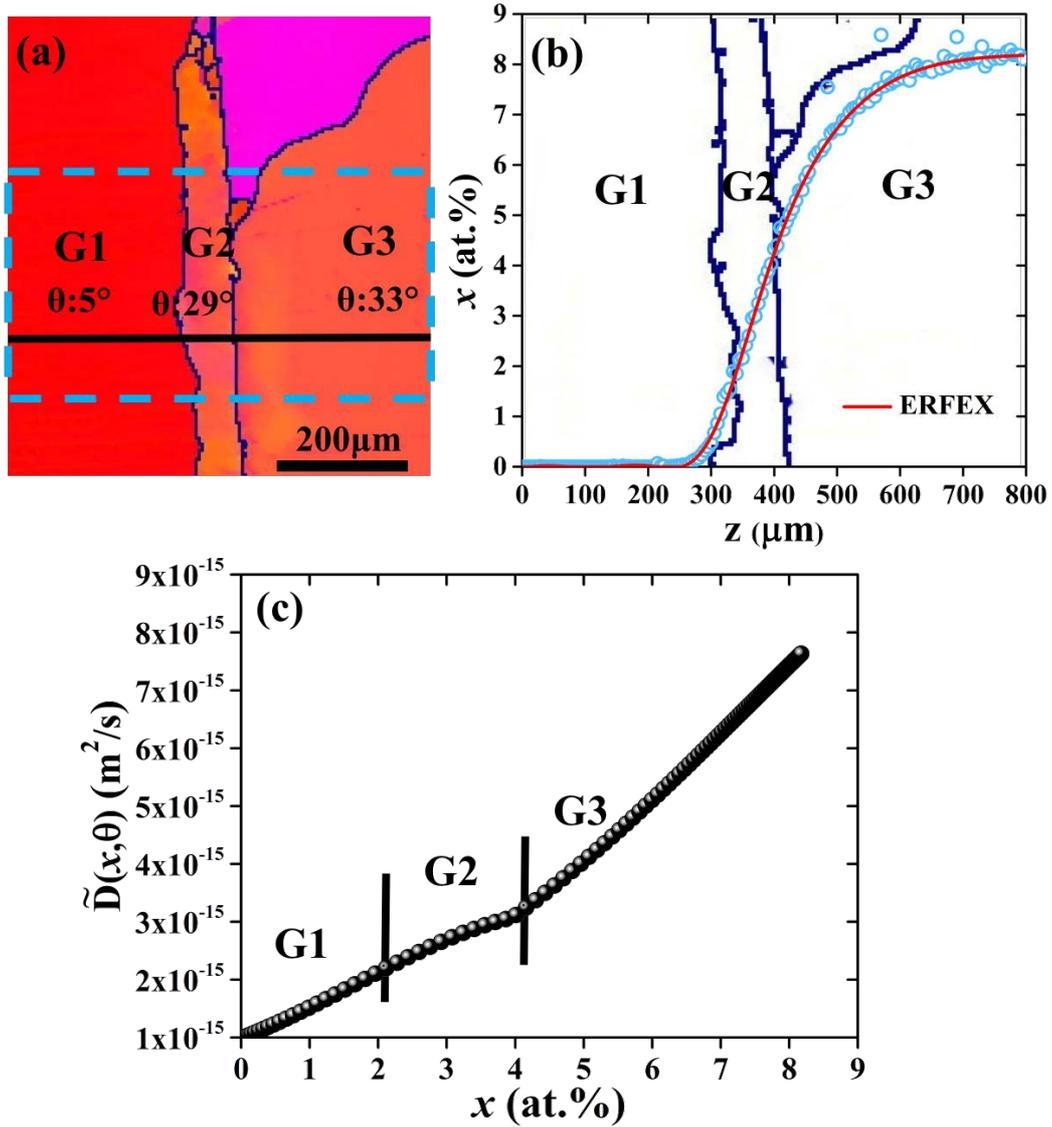

Fig. 5 (a) Microstructure in the diffusion zone where the color code indicates the grain orientation according to the inverse pole figure. (b) The experimental composition profile (blue circles) measured by EPMA along the black line in (a), combined with the analytical fit using the ERFEX error expansion function. (c) Extracted interdiffusion coefficients as a function of the Al content for each grain.

Since the diffusion profile inevitably passes multiple grains with diverse grain orientation, the interdiffusion coefficients along the composition profile also depend on the grain orientation. The interdiffusion coefficients in each grain, $\widetilde{D}(x,\theta)$, were obtained using the forward-simulation scheme originally developed by Zhang and Zhao [33] for multi-phase diffusion. To do this, the interdiffusion coefficients were firstly determined using the traditional Boltzmann-Matano [34] method or the Sauer-Freise method [35] based on the composition profiles. $\widetilde{D}(x,\theta)$ was then expressed by a quadratic function of the Al content (in at. %) in each grain according to

$$Log_{10}\widetilde{D}(x,\theta_i) = p_i x^2 + q_i x + r_i, \tag{2}$$

where $p_i$, $q_i$ and $r_i$ stand for the fitting parameters, which vary with the grain orientation. Afterwards, the composition profile in each grain was simulated by solving Fick's second law of diffusion in a form of ordinary differential equation transformed by the Boltzmann transformation

$$-\frac{\xi}{2}\frac{dx}{d\xi} = \frac{d}{d\xi}\left(\widetilde{D}(x,\theta_i)\frac{dx}{d\xi}\right). \tag{3}$$

Finally, the fitting parameters in the eq. (2) were adjusted to fit the experimental composition profile for each grain in Fig. 5b and the interdiffusion coefficient was obtained. Note that continuity of the diffusion flux across grain boundaries was assumed because the grain boundary width is negligibly small compared to the typical diffusion distance (~ mm). The interdiffusion coefficients for grains G1, G2 and G3 are plotted in Fig. 5c as function of the Al content. As expected, the variation of the interdiffusion coefficients with the Al content differs in each grain.

### 3.2 Splitting $D_\perp$ and $D_\parallel$ along the basal plane and parallel to the c-axis

After the interdiffusion coefficients were obtained as a function of the grain orientation and composition $\widetilde{D}(x,\theta)$, the anisotropic interdiffusion coefficients $D_\perp$ and $D_\parallel$ were extracted using the theoretical framework proposed by Shewmon [21], which assumed that

$$\widetilde{D}(x,\theta) = D_\perp(x)sin^2(\theta) + D_\parallel(x)cos^2(\theta). \tag{4}$$

$D_\perp(x)$ and $D_\parallel(x)$ could be expressed using an Arrhenius law, leading to:

$$D_\perp(x) = D_{\perp 0}(x)exp\left(-\frac{Q_\perp(x)}{RT}\right) \tag{5}$$

$$D_\parallel(x) = D_{\parallel 0}(x)exp\left(-\frac{Q_\parallel(x)}{RT}\right) \tag{6}$$

where $D_{\perp 0}(x)$ and $Q_{\perp}(x)$ are the frequency factor and diffusion activation energy within the basal plane, whilst $D_{\parallel 0}(x)$ and $Q_{\parallel}(x)$ stand for the corresponding values parallel to the $c$ axis of the grain. In the present work, it was assumed that both the activation energy and the frequency factor have a linear dependence with the composition according to

$$D_{\perp 0}(x) = a_{\perp}x + b_{\perp} \qquad D_{\parallel 0}(x) = a_{\parallel}x + b_{\parallel} \qquad (7)$$

$$Q_{\perp}(x) = c_{\perp}x + d_{\perp} \qquad Q_{\parallel}(x) = c_{\parallel}x + d_{\parallel} \qquad (8)$$

where $a_{\perp}$, $b_{\perp}$, $c_{\perp}$, and $d_{\perp}$ (and $a_{\parallel}$, $b_{\parallel}$, $c_{\parallel}$, and $d_{\parallel}$) are the adjustable parameters. In case of more complex compositional dependency, a quadratic function could be assumed as well. They were obtained from the experimental values of $\widetilde{D}(x,\theta)$ through eq. (4) using the least squares fitting and, then, $D_{\perp}$ and $D_{\parallel}$ can be determined.

### 3.3 Effect of grain boundary on interdiffusion

In practice, it is hardly to obtain a diffusion couple that includes a single grain across the diffusion region because of the recrystallization along the diffusion interface during the high temperature annealing, as shown in Fig. 4. As the diffusion rate of the Al atoms along the grain boundaries is known to be two orders magnitude faster than that in bulk Mg [36], it is critical to eliminate any grain boundary diffusion contribution in the regions of the diffusion couple used to determine the diffusion coefficients.

The effect of grain boundaries on the composition profile is shown in Fig. 6 for two typical cases found in the diffusion couples. In the first case (Fig. 6a), the composition profile was measured along the line S1 (red) which crossed three grains denoted A, B, and C. The grain boundaries A/B and B/C were perpendicular to the composition line S1. In the second case (Fig. 6b), a small grain F was found between the large grains E and G. The composition profile was measured along two lines, the solid line S2a (dark) and dashed line S2b (blue). Both the lines were perpendicular to the grain boundary E/G, but the line S2a was located close to (around 40 μm away) the grain boundary E/F, which was parallel to the line S2a.

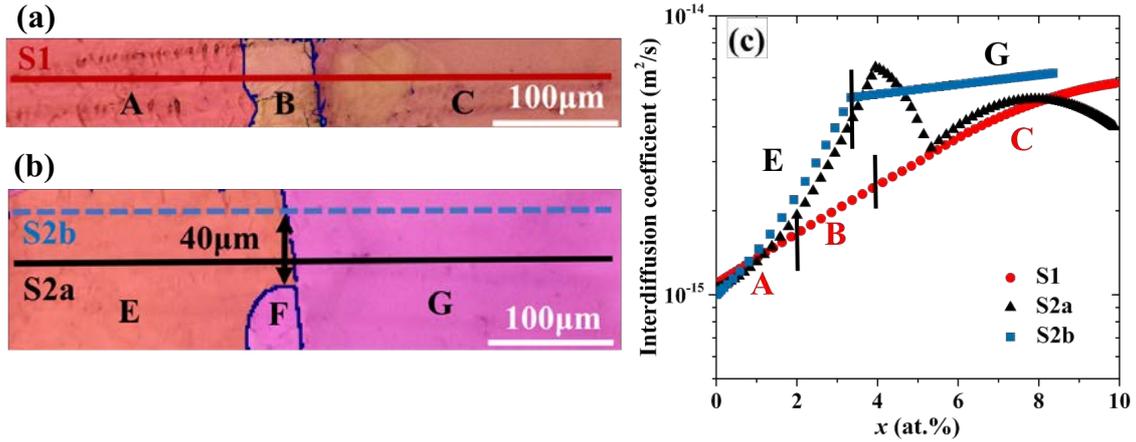

Fig. 6 Microstructure of the diffusion couple near the diffusion interface of the two typical cases. The colour code indicates the grain orientation (according to the IPF). Grain boundaries are delineated by the blue line. (c) Interdiffusion coefficients extracted from the composition profiles along the lines S1 and S2a and S2b.

The interdiffusion coefficients extracted from the composition profiles along the three lines in Figs. 6a and 6b using the approach indicated above are plotted as a function of the Al content in Fig. 6c. Apparently, the orientation differs in each grain, and hence the interdiffusion coefficients were different within each grain. There was not kink at the grain boundaries in the case of the lines S1 and S2b, because these lines were perpendicular to the grain boundaries and the grain boundary leakage (perpendicular to the diffusion direction) showed no essential influence. Nevertheless, the interdiffusion coefficients increased by a factor of 7 along the line S2a in the region close the grain boundary E/F. It should be noted that they were mainly found in the Mg-Al part of the diffusion couple (Fig. 4) because recrystallization occurred often during annealing in the soft Mg region of the diffusion couple, leading to the development of small grains. As a consequence, only composition profiles far away (> 100 μm) of any lateral grain boundary in Fig. 4 were utilized to determine the interdiffusion coefficients to avoid the effect of grain boundary leakage.

## 4. Results and discussion
### 4.1 Influence of grain orientation on the interdiffusion coefficients

Composition profiles perpendicular to the diffusion interface were obtained along scan lines such that any grain boundary and its effect close to the scan line (such as the S1 scan line in Fig. 6a) can be avoided. The interdiffusion coefficients for grains with different orientations were extracted by the forward-simulation scheme described in Section 3 and the $p_i$, $q_i$ and $r_i$ coefficients that account for the dependence of the Al content and of the rotation angle θ on the interdiffusion coefficients for are given in Table 1. The interdiffusion coefficients $\widetilde{D}(x, \theta_i)$ were assumed to be expressed by a quadratic function of the Al content (in at. %) in each grain using eq. (2).

Table 1. Coefficients of eq. (2) to express the interdiffusion coefficients as a function of Al content for different rotation angles θ.

| Temp (K) | Rotation angle | $p_i$ | $q_i$ | $r_i$ |
|---|---|---|---|---|
| 673 K | 5.8° | -362.8237 | 61.2367 | -35.5152 |
| | 33° | -128.4483 | 32.9941 | -34.7475 |
| | 44° | 53.9833 | -1.4807 | -33.0104 |
| | 75° | 34.7870 | 1.5273 | -33.0756 |
| 723 K | 5.3° | 266.6762 | -19.9245 | -31.1241 |
| | 28° | 450.7663 | -33.9376 | -31.1299 |
| | 53° | 614.2203 | -47.3864 | -30.6317 |
| | 81° | 320.2773 | -41.5305 | -29.7505 |

The influence of the grain orientation (described by θ) on the interdiffusion coefficients is shown in Fig. 7. The results in Fig. 7a summarize the interdiffusion coefficients from Mg-5 at.%Al to Mg-8.5 at.%Al at 673 K whereas those in Fig. 7b cover the composition from Mg-5.5 at.%Al to Mg-7.5 at.% Al at 723 K. These results indicate that the interdiffusion coefficients increased with the Al content as well as with θ, implying that diffusion along the basal plane was faster than along the *c* axis. Similar trends were found for the Al and Zn impurity diffusion coefficients in Mg [25]. Nevertheless, a minimum in the diffusion coefficients was found for θ ~30°, as shown in Fig. 7b, and this minimum was present for all the different Al contents. The minima are being further scrutinized by our ongoing experimental investigation and theoretical treatment.

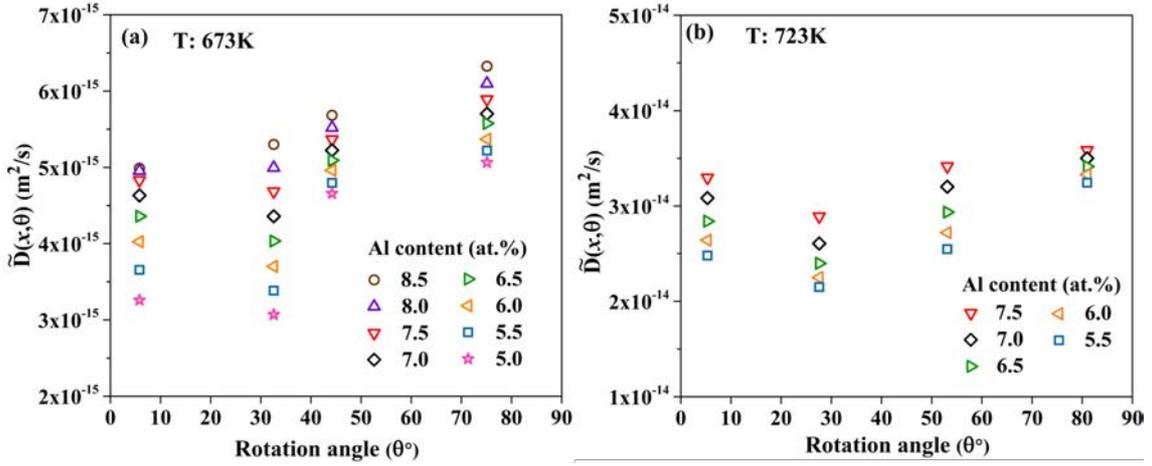

Fig. 7 Influence of the rotation angle θ on the interdiffusion coefficients in Mg-Al alloys. (a) From Mg-5 at.%Al to Mg-8.5 at.%Al at 673 K. (b) From Mg-5.5 at.%Al to Mg-7.5 at.%Al at 723 K.

The two interdiffusion coefficients $D_\perp(x)$ and $D_\parallel(x)$ in the basal plane and along the $c$ axis were obtained from the interdiffusion coefficients following the splitting scheme described in Section 3.2 and they are plotted in Fig. 8. Moreover, the parameters in eqs. (7) and (8) to obtain the interdiffusion coefficients as a function of the Al content and temperature are listed in Table 2. As expected, $D_\perp(x)$ and $D_\parallel(x)$ increased with the Al content and with temperature. Note that the interdiffusion along the basal plane was constantly faster than along the $c$ axis for the same Al content and temperature, and that the differences decreased as the Al content increased.

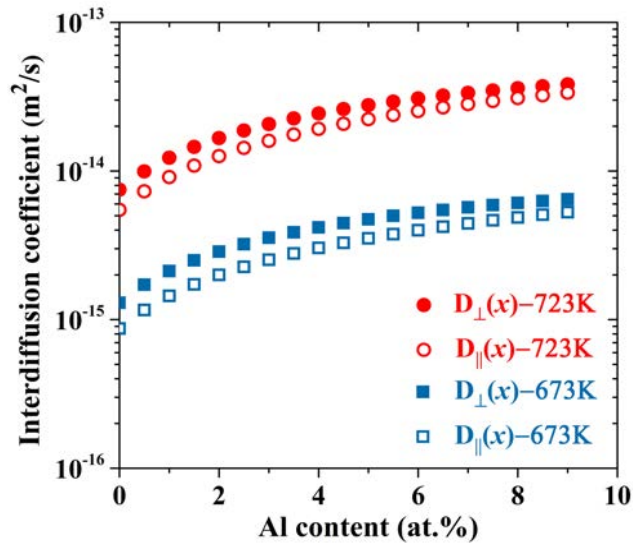

Fig. 8 $D_\perp(x)$ and $D_\parallel(x)$ interdiffusion coefficients as a function of the Al content at 673 K and 723 K.

The accuracy of the interdiffusion coefficients was confirmed by the comparison between the experimental and the calculated interdiffusion coefficients in the present work, shown in Fig. 9. The dashed lines refer to the generally accepted deviation range of a factor between 0.5 and 2. A good accordance between the calculated values and experimental data is achieved in this investigation. The errors resulted from the determination of anisotropic interdiffusion coefficients can be analysed by a robust treatment proposed by Lechelle.et al [37], wherein the error propagation was considered in details. The method has been successfully applied to the diffusion study in many ternary systems [38,39]. Because the main purpose of this work is to present and demonstrate a high throughput methodology to extract the anisotropic interdiffusion coefficients, a simplified protocol was utilized to analyse the errors in the extracted interdiffusion coefficients $D_\perp(x)$ and $D_\parallel(x)$. By mainly considering the smoothing of the raw composition profile and the splitting the interdiffusion coefficients, the errors were estimated to be approximately 8%.

Moreover, the anisotropic interdiffusion coefficient in Mg-Al alloy $\widetilde{D}(x,\theta)$ could be calculated based on eq. (4) as a function of the Al content and grain orientation at 673 K and 723 K, and are plotted in Figs. 10a and b for 673K and 723K, respectively. The plot, superimposed with experimental results obtained in this work, graphically illustrates the influence of Al content and grain orientation on the interdiffusion diffusion in the hcp lattice.

Table 2. Parameters to derive the diffusion activation energy and frequency factor of the interdiffusion coefficients along the basal plane ($D_\perp$) and the $c$ axis ($D_\parallel$,) as defined by eqs. (7) and (8).

| $D_{0\perp}$ (m$^2$/s) | | $Q_\perp$ (J/mol) | | $D_{0\parallel}$ (m$^2$/s) | | $Q_\parallel$ (J/mol) | |
|---|---|---|---|---|---|---|---|
| $a_\perp$ | $b_\perp$ | $c_\perp$ | $d_\perp$ | $a_\parallel$ | $b_\parallel$ | $c_\parallel$ | $d_\parallel$ |
| 9.42 10$^{-3}$ | 1.33 10$^{-4}$ | 24.2 10$^3$ | 1.42 10$^5$ | 2.10 10$^{-2}$ | 3.03 10$^{-4}$ | 1.10 10$^4$ | 1.49 10$^5$ |

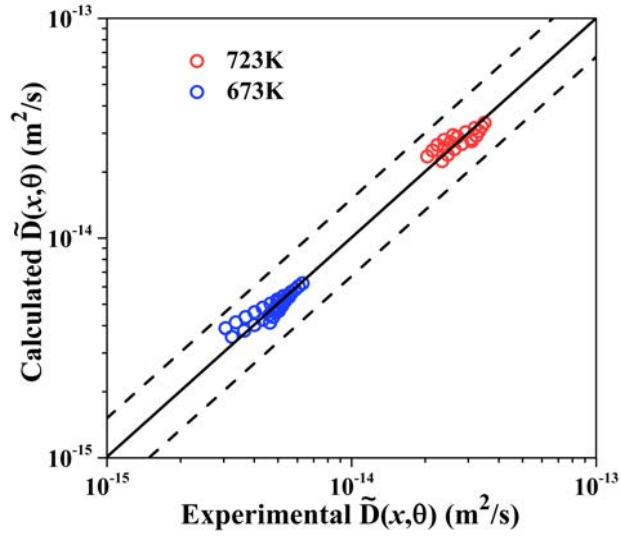

Fig. 9 Comparison between the experimental and calculated anisotropic interdiffusion coefficients of Mg-Al alloy at 673K and 723K. The dashed lines indicate that the differences between experimental and calculate values are within the range of a factor between 0.5 and 2.

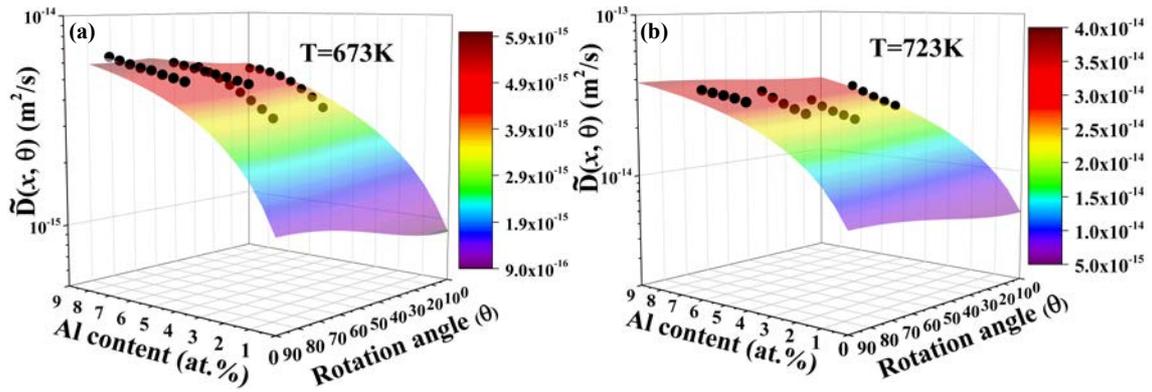

Fig. 10 Calculated interdiffusion coefficients as a function of the Al content and grain orientation according to eq. (4). (a) 673K. (b) 723K. Black circles stand for the experimental values obtained in the present work.

## 4.2 Anisotropic impurity diffusion coefficients

The Al impurity diffusion coefficients within the basal plane, $D^{Mg}_{Al\ \perp}$, and along the $c$ axis of the grain, $D^{Mg}_{Al\ \parallel}$, could be extrapolated from the two interdiffusion coefficients along $D_\perp(x)$ and $D_\parallel(x)$ as the Al content approaches zero. The Al impurity diffusion coefficients in hcp Mg extrapolated from the results in this work are plotted as function of 1000/T (following the Arrhenius law) in Fig. 11a, together with the experimental impurity diffusion coefficients obtained from Mg polycrystals [13–15,25,40] and single crystals along the basal plane and the $c$ axis [27]. The present results are consistent with the experimental values obtained by Brennan [15] and Kammerer [13] using polycrystalline Mg-Al diffusion-couples and also with the first principles calculations of Zhou [29]. Nevertheless, they were much smaller than those measured by Brennan [14] using the secondary ion mass spectrometry in polycrystalline Mg and by Das [27] using the diffusion couples in the single crystals. In fact, Brennan [14] already mentioned that the surface roughness introduced from the sputtering process could result in the artificial overestimation of the diffusion coefficients, which led to unreasonable higher values. Generally, these experimental values measured from the polycrystalline Mg are expected to be of mixed diffusion between the volume diffusion (itself varies with the distinct grain orientations) and grain boundary diffusion, i.e. a sort of average diffusion coefficient. The experimental anisotropic impurity diffusion coefficients were experimentally measured by Das [27] within the basal plane and along the *c*-axis that were even much higher than any of the previous values in the polycrystals [13–15], as shown in Fig. 11(a), which are considered to be controversial and not taken into account in the present work.

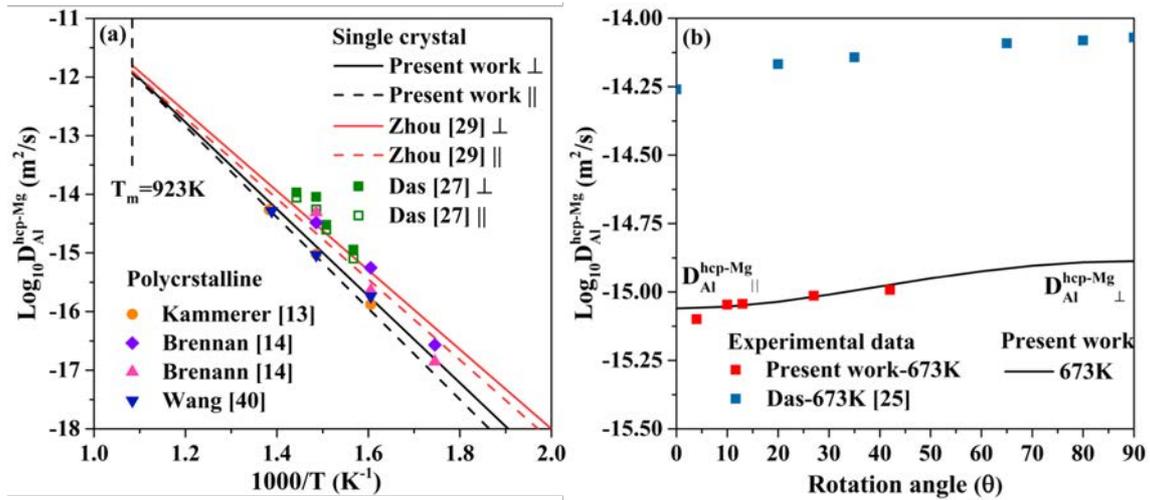

Fig. 11 (a) The Arrhenius expression of the Al impurity diffusion coefficient in pure Mg. The different symbols stands for experimental results in the literature for polycrystals [13–15,27,40] and single crystals [27]. The solid lines stand for the predicted Al impurity diffusion coefficients perpendicular to the *c* axis, while the dashed lines represent the predictions along the c axis. Predictions from this investigation are shown as black lines while first-principles calculations are plotted as red lines [29]. (b) Influence of the rotation angle (θ) on the Al impurity diffusion coefficient at 673 K (black line), along with the experimental data from the present work (red squares) and Das *et al.* at 673 K (blue squares) [25].

The Al impurity diffusion coefficients as a function of the grain orientation in pure hcp-Mg were successfully extrapolated at the dilute end of the Al content from the diffusion composition profiles. They are plotted in Fig. 11b, together with the computed values based on eq. (4) when $x \to 0$. The experimental data from Das [25] at 673 K are also plotted in Fig. 11b for comparison. Those data are almost one order of magnitude higher and, as discussed above, the present results are more reliable and reasonable.

### 4.3 Anisotropy ratio of the interdiffusion coefficients

The anisotropy ratio $D_\perp(x)/D_\parallel(x)$ between the interdiffusion coefficients along the basal plane and the $c$ axis is widely used as an indicator of the anisotropy in diffusion process of the hcp materials. Regarding the self-diffusion coefficients of hcp materials, the anisotropy ratio correlates well with the $c/a$ ratio of the lattice parameters. Experimental results of the self-diffusion anisotropy ratio show that $D_\perp/D_\parallel > 1$ in α-Ti [41], α-Zr [42], α-Hf [43] and Mg [23,24] with $c/a < \sqrt{8/3}$, whereas $D_\perp/D_\parallel < 1$ in Zn [44–46] and Cd [47] with $c/a > \sqrt{8/3}$. As for the self-diffusion hcp-Mg [23,24], the anisotropy ratio of the self-diffusion of the hcp Mg was 1.13 at 741 K and increased to 1.24 at 848 K [23], while $c/a = 1.6236 < \sqrt{8/3}$ [48], in good agreement with previous observations.

Regarding the impurity diffusion behaviour, the anisotropy ratio $D_\perp(x)/D_\parallel(x)$ was found to be < 1 for some solute atoms (like Al, Zn, Y, and Gd [18,26,27,29,49]). Among them, the anisotropy ratio of the Al impurity diffusion coefficients was about 1.3 and decreased as the temperature increased [27]. All results support the fact that the self-diffusion and impurity diffusion are faster within the basal plane.

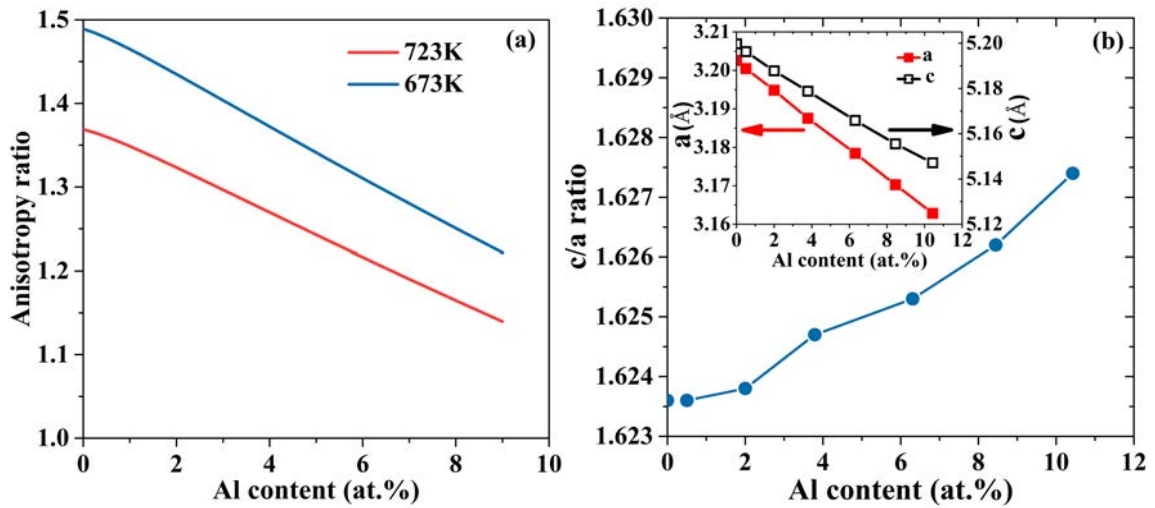

Fig. 12 (a) Evolution of the interdiffusion anisotropy ratio as a function of the Al content at 673 K and 723 K. (b) Evolution of the $c/a$ ratio in hcp Mg with the Al content [48]. The variation of the lattice parameters $a$ and $c$ of hcp Mg with the Al content are shown in the inset.

The extraction of the anisotropic interdiffusion coefficients in this work also allows to evaluate the interdiffusion anisotropy ratio $D_\perp(x)/D_\parallel(x)$ as the function of the Al content in the hcp Mg-Al binary alloy. The variation of interdiffusion anisotropy ratio in hcp-Mg with the Al content at 673 K and 723 K is plotted in Fig. 12a, showing $D_\perp/D_\parallel > 1$ over the entire Mg-Al composition range. Thus, the interdiffusion is faster within the basal plane than that along the $c$ axis regardless of the Al content, however, the differences decrease as the Al content or the temperature increased, in agreement with the trends reported for the self-diffusion of pure Mg and Al impurity diffusion in Mg. The ratio extrapolated to the infinite dilute end of Al in hcp-Mg is about 1.48 at 673 K and ~ 1.37 at 723 K, similar to the reported anisotropy ratio 1.3 of the impurity diffusion of Al in hcp Mg [27]. The decrease of the interdiffusion anisotropy ratio is inherently associated with the change of the lattice parameters a and c (and thus, the *c*/*a* ratio) with the Al content in the Mg-Al binary alloy [48], as depicted in Fig. 12b. The increase of the Al content continuously reduced a and c, because Al has a negative size misfit with respect to the Mg host. But this reduction is larger for a in the basal plane, leading to an increase in the *c*/*a* ratio with the Al content, gradually approaching the ideal *c*/*a* = 1.633. Therefore, the addition of Al solutes had a larger effect in the basal plane, leading to the larger solute-vacancy exchange barrier along the basal plane and reducing the interdiffusion component $D_\perp(x)$ within the basal plane, as compared to $D_\parallel(x)$ along the $c$ axis. The explanation of the change in the anisotropy of diffusion with the Al content follows from a consideration of the difference in the saddle points due to the introduction of Al in the hcp Mg lattice, this updates the binding energy for diffusion in and out of the basal plane and the $c$ axis [50]. Theoretically, the change of the configuration of the lattice by the increase of Al content in Mg-Al alloys also plays a key role in the formation vacancy energy in and out of basal plane, which requires further investigation.

## 5. Conclusions

By combining information of the composition profiles obtained with the EPMA analysis and the grain orientation spectrum with the EBSD test in polycrystalline diffusion couples, a high-throughput experimental approach was presented to investigate the anisotropy of the diffusion behavior of hcp crystals. The interdiffusion coefficients of Mg-Al alloy were extracted at 673 K and 723 K as a function of the grain orientation and Al content. The leakage of grain boundaries in the polycrystalline system was analyzed and was found to be negligible when the grain boundaries were perpendicular to the scanning route of the composition profiles. The minimum lateral distance between the scan lines and the grain boundaries to avoid grain boundary leakage was found to be 40 µm in the Mg-Al alloy.

The anisotropic interdiffusion coefficients $D_\perp(x)$ and $D_\parallel(x)$, perpendicular and parallel to the *c* axis of the hcp Mg lattice, were determined as a function of the Al solute content from the experimental interdiffusion coefficients in grains with different orientation. $D_\perp(x)$ and $D_\parallel(x)$ increased with the Al content and the interdiffusion coefficients within the basal plane were 1.25 times and 1.15 higher than those along the *c* axis at 673 K and 723 K, respectively. The anisotropy factor $D_\perp(x)/D_\parallel(x)$ decreased with the Al content at both 673 K and 723 K. Finally, an explicit expression of the interdiffusion coefficients as a function of Al content, grain orientation and temperature was derived. This information can be used as input in models of microstructure formation and solute redistribution resulting from such processes as solidification and precipitation, together with the crystallographic or thermodynamic anisotropy.

**Acknowledgments**

This work was supported by the Natural Science Funds of China [Grant no. 51571113] and by the European Research Council (ERC) under the European Union's Horizon 2020 research and innovation programme (Advanced Grant VIRMETAL, grant Agreement no. 669141). Ms. J-Y. Wang acknowledges the financial support from the China Scholarship Council (Grant no. 201506890002). GX acknowledges the support from the Natural Science Funds of China [Grant no. 51701094]. The authors are thankful to Prof. Xiaoma Tao at Guangxi University for the EPMA measurement.